\documentclass[amsmath,amssymb,prl,twocolumn,a4paper]{revtex4}
\usepackage{graphicx}
\usepackage{dcolumn}
\usepackage{bm}
\usepackage{amsmath}
\usepackage{amssymb}
\usepackage{epic,eepic}
\usepackage{subfigure}
\usepackage{overpic}
\begin{document}

\author{Davide Cellai}
\affiliation{
Irish Centre for Colloids and Biomaterials, Dept. of
Chemistry, University College Dublin, Belfield, Dublin 4}

 \author{Hector Cuevas}
\affiliation{
Irish Centre for Colloids and Biomaterials, Dept. of
Chemistry, University College Dublin, Belfield, Dublin 4}

\author{Aonghus Lawlor}
\affiliation{
Irish Centre for Colloids and Biomaterials, Dept. of
Chemistry, University College Dublin, Belfield, Dublin 4}

\author{Gavin D. McCullagh}
\affiliation{
Irish Centre for Colloids and Biomaterials, Dept. of
Chemistry, University College Dublin, Belfield, Dublin 4}

\author{Kenneth A. Dawson}
\affiliation{
Irish Centre for Colloids and Biomaterials, Dept. of
Chemistry, University College Dublin, Belfield, Dublin 4}

\title{ Competition between Short-Ranged Attraction and
Short-Ranged Repulsion in
Crowded Configurational Space; A Lattice Model Description}

\begin{abstract}
We describe a simple nearest-neighbor Ising model that is capable
of supporting a gas, liquid, crystal, in characteristic relationship
to each other. As the parameters of the model are varied one
obtains characteristic patterns of phase behavior reminiscent of
continuum systems where the range of the interaction is varied.
The model also possesses dynamical arrest, and although we have
not studied it in detail these 'transitions' appear to have a
reasonable relationship to the phases and their transitions.

\end{abstract}

\date{\today}
\maketitle

In many systems, one can observe a rich interplay between
phase-separation, critical phenomena, slowed dynamics and
solidification, the latter manifesting itself as both
crystallization and glassification. These questions have a long
and venerable past, but in modern condensed matter theory the
range of systems where the issues have become relevant
\cite{dawsoncurrop,relevant,bergenholtz} is remarkable.
Recent interest in
model systems extend from colloidal glasses \cite{colloid},
particle gels
\cite{bergenholtz,gels}, polymeric gels
\cite{polymericgels}, globular protein
crystallization
\cite{crystallization,foffipinidawson}
and
gellation. For example, it has recently transpired that even the
simplest system with repulsive core and short ranged attraction
exhibits a large range of phase-transitions, dynamical arrest
(possibly with new dynamical logarithmic singularity
\cite{gels,singularity}) and kinetic phenomena
\cite{phenomena},
the pertinent control parameter being the range of the attraction.
So far it has not been possible to understand the inter-relation
of all these effects.
Such
issues also lie at the heart of many important modern problems of
materials science, including the formation of arrays of particles
on optical wavelengths and knowledge-based materials design
\cite{frenkelscience,lekkerkerkernature}.

To fully explore these questions we would need to describe an
extended range of density, across a number of different phase
boundaries, dealing naturally with criticality
\cite{criticality}, metastabiliy, and
arrest phenomena \cite{rosenbaum},
all in a coherent fashion, with tools that were
applicable and reliable across these regimes.
In the interim progress
can be made in various aspects of the problem
\cite{frenkelscience,foffipinidawson,caccamo}.

The idea that the gas, liquid, crystal, and transitions between
them can be studied within the same lattice model has been raised
over the years in some interesting studies \cite{studies}, and
complex models have also been studied using such simple models \cite{approaches}.
However, there has never been any simple (lattice-based) general
mechanism producing a liquid (considered as a large collections of
attraction-dominated states with nearly-degenerate energies), in
appropriate relation to gas and solid.
Here, beginning from ideas introduced by Biroli and M\'ezard
\cite{birolimezard,lawlorprl} we show for the first time that it
is possible to caricature all the states, and their transitions,
in a remarkably simple nearest-neighbor Ising model. In this model
dynamical arrest and glassy states are also naturally incorporated
into the story.

In the model, space is divided into cubes of side $a$,
characteristic of the particle size, and the microscopic length of
the system. To the center of each cube we associate a site
(coordination number $c=6$, in a tridimensional space)
and Ising-like occupancy variable. Where a site is occupied we
define (respectively) attractive and repulsive interactions
between $c_a$ and $c-c_r$ of the nearest neighbors.

Thus $c_r$ specifies the fraction of space within a given cube
that is available to neighboring particles, given the nature of
the repulsive interactions, whilst $c_a$ specifies how many of
those particles can benefit from the attractive energy of the
system. Whilst we emphasize that this is a caricature, rather than
a representation, of the potential, we expect that $c_r$ therefore
expresses the complexity or irregularity of the core of the
particle, reflected ultimately in the density at random close
packing in the real system, and in the model. The interpretation
of $c_a$ is more subtle. However, broadly speaking, a small number of
($c_a$) attractive interactions are shared between a larger number
($c_r$) neighbors, there are many local configurations in which
the attractive contribution saturates in
the dense system. This freedom encourages the formation of a
liquid. This idea is consistent with the rationale for formation
of a gel by short-ranged colloidal particles \cite{dawsoncurrop}.
In that case there is such limited freedom by which the
attractions can be captured that the liquid arrests to form a
gel-like state. There is also a general trend for the model to be
dominated by the 'liquid' state as we approach $c_r=c_a=6$, the
nearest-neighbor model.

In our present study, we assume nearest-neighbor finite
attractions ($\epsilon$) and infinite repulsions, the latter to
make contact with related studies \cite{birolimezard,lawlorprl}.

Thus, the infinite repulsion between $c_r$ neighbors may be
written ($0\leq c_r \leq 6$):
\begin{equation}
  E_j^{\textrm{repulsive}}=\left\{
  \begin{array}{ll}
   0 & \textrm{if $l_j \leq c_r$}\\
   +\infty     & \textrm{otherwise}\\
  \end{array}
  \right.
\end{equation}
where $l_j$ is the number of occupied nearest-neighbor sites. In
the case of $c_r=6$ no repulsive interaction is present in the
model. The attractive interaction is governed by the parameter
$c_a$ ($0 \leq c_a \leq 6$) and is defined (per particle) as
\begin{equation}
  E_j^{\textrm{attractive}}=\left\{
  \begin{array}{ll}
   -l_j\epsilon & \textrm{$l_j\leq c_a$}\\
   -c_a\epsilon     & \textrm{otherwise}\\
  \end{array}
  \right.
\end{equation}
\noindent where $\epsilon>0$ is the strength of the attraction.
Throughout the paper we will define energies and chemical
potential in units of $k_bT$, and the 'effective' energy will refer
to the combined exponent of energy and chemical potential in the
partition function.

Despite the fact that this is a simple nearest-neighbor Ising
model, it is remarkably rich and spatial frustration produces a
large range of phenomena, some of which have previously been
observed in more complicated lattice models. In such cases it has
also been possible to formulate methods for their study
\cite{approaches,fisher,study,dawsonpr}. For the present case we
show that the existence of a liquid disposed between the gas and
liquid, and resulting triple point originate in a large degeneracy
of states that are found at zero temperature as 'multiphase
points'. In special cases we make the number of such states
macroscopic, but even where they are not so, at finite
temperatures we can cause the existence of an infinite number of
near-degenerate states, and a liquid.

It is possible to determine the phase-diagram of such models to a
high degree of confidence. To begin with we determine the ordered
phases by construction of the zero temperature states of the
model. On the simple cubic lattice we re-write the underlying
effective Hamiltonian (itself a combination of pure energy and
chemical potential) as a sum over octahedral fragments
\cite{lipkin,dawsonpr} to each of which we can associate an
independent energy. We may then choose the 'optimal' fragments that
are able to tile the lattice, and thereby classify the
zero-temperature states.
There is at least one
multiphase point \cite{study,dawsonpr,lipkin,sinai} at
which an infinity of zero-temperature states is degenerate, though
the entropy per particle remains finite.

In general terms the zero temperature states are organized as
follows. In three dimensions, there are $20$ different types of
fragment.
Now, in the Grand Canonical ensemble, for fixed values
of $c_r$ and $c_a$, the energy of a particle-centered fragment has
the form $-q\epsilon-\mu$, where $\mu$ is the chemical potential
and $q$ is a positive integer number depending on the number of
neighbors. All the vacancy-centered fragments have zero energy,
unlike previous frustrated models. Therefore we use a slightly
different notation, defining as $n$ and $a$ the total number of
nearest neighbors and the number of filled axes (linear triplets
of sites, the extreme pair being occupied), respectively, and
using a label $p=0,1$ on the left of the symbol of a fragment
distinguishing the vacancy-centered and the particle-centered
ones: ${}^{\phantom{a}}_pf_n^a$.

\begin{figure}[hbp]
 \begin{center}
 \includegraphics[width=\columnwidth]{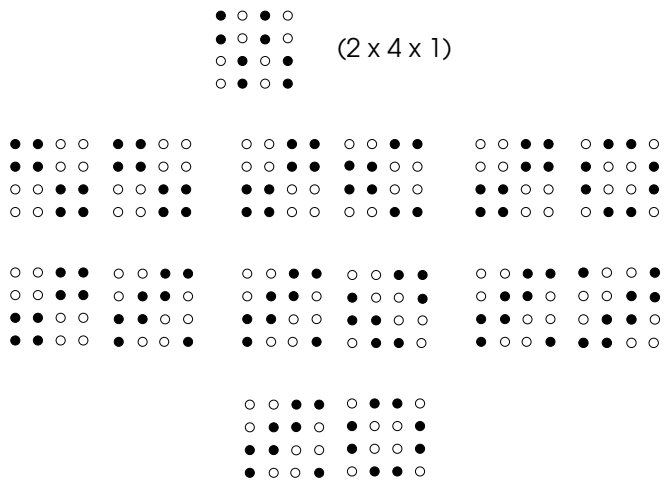}
 \caption{Pure states for the case $c_r=3,4$, $c_a=3$.
 The first is made up of $f_3^1$ fragments (top).
 There are seven pure phases made from
 the $f_3^0$ fragments.
 Each of the seven pairs constitues  two layers  of a
 $4\times 4\times 4$ unit cell the remainder being
 the particle-vacancy symmetric ones.}
 \label{fig:c3c3pure}
 \end{center}
\end{figure}

The internal energy per volume of a state of the model
is given by:
\begin{equation}
 \frac{U}{L^3}=-\left(\sum_j^{PC}q_j\rho_j\right)\epsilon
    -\left(\sum_j^{PC}\rho_j\right)\mu ,
\end{equation}

\noindent
where $L^3$ is the lattice volume and $\rho_j$ is the density of
the $j-$th fragment; the sum is intended only over the
particle-centered fragment types (PC).

For $\mu\to -\infty$ the ground state is the zero-density gas,
because in that limit the energy of every particle-centered
fragment is infinite.

For $\mu\to +\infty$ the ground state is the most dense crystal,
i.e. the crystal with the lowest density of vacancy-centered
fragments. In between we have many different situations,
depending on the choice of $c_a$ and $c_r$. Below we give only
illustrative examples.

Thus, for the case $c_a=3$ and $c_r=3$ we have two fragments with
energy $-3\epsilon-\mu$, but only one (${}^{\phantom{0}}_1f_3^0$)
tiles the space with the highest density vacancy-centered crystal
${}^{\phantom{0}}_0f_6^3$. So there is one highest density crystal
which is ordered along the diagonal with a repeat pattern of an
occupied pair of layers and an unoccupied layer. This state is
degenerate with the zero-density gas at the point
$\mu=-3\epsilon$,
 the zero-temperature boundary between gas and the crystal.
There cannot be finite regions occupied by other states at zero
temperature. In fact, a different state could have lower energy
only if
\begin{equation}
 q_{PC}=\frac{\sum^{PC}_{j}q_j\rho_j}{\sum^{PC}_{j}\rho_j}>3 .
\end{equation}
and this is impossible because $q_j\leq 3$ for every $j$.\\
The point $\mu=-3\epsilon$ is a multiphase point, because
two particle-centered fragments (${}^{\phantom{0}}_1f_3^0$,
${}^{\phantom{0}}_1f_3^1$) and
every vacancy-centered fragment is degenerate.
Some examples of crystals at the multiphase point
are given in Figure \ref{fig:c3c3pure}.

\begin{figure}[!htp]
\begin{center}
 \subfigure[]{
  \label{fig:subfig:crysthector}
  \includegraphics[width=0.40\columnwidth,angle=0,keepaspectratio=T]
  {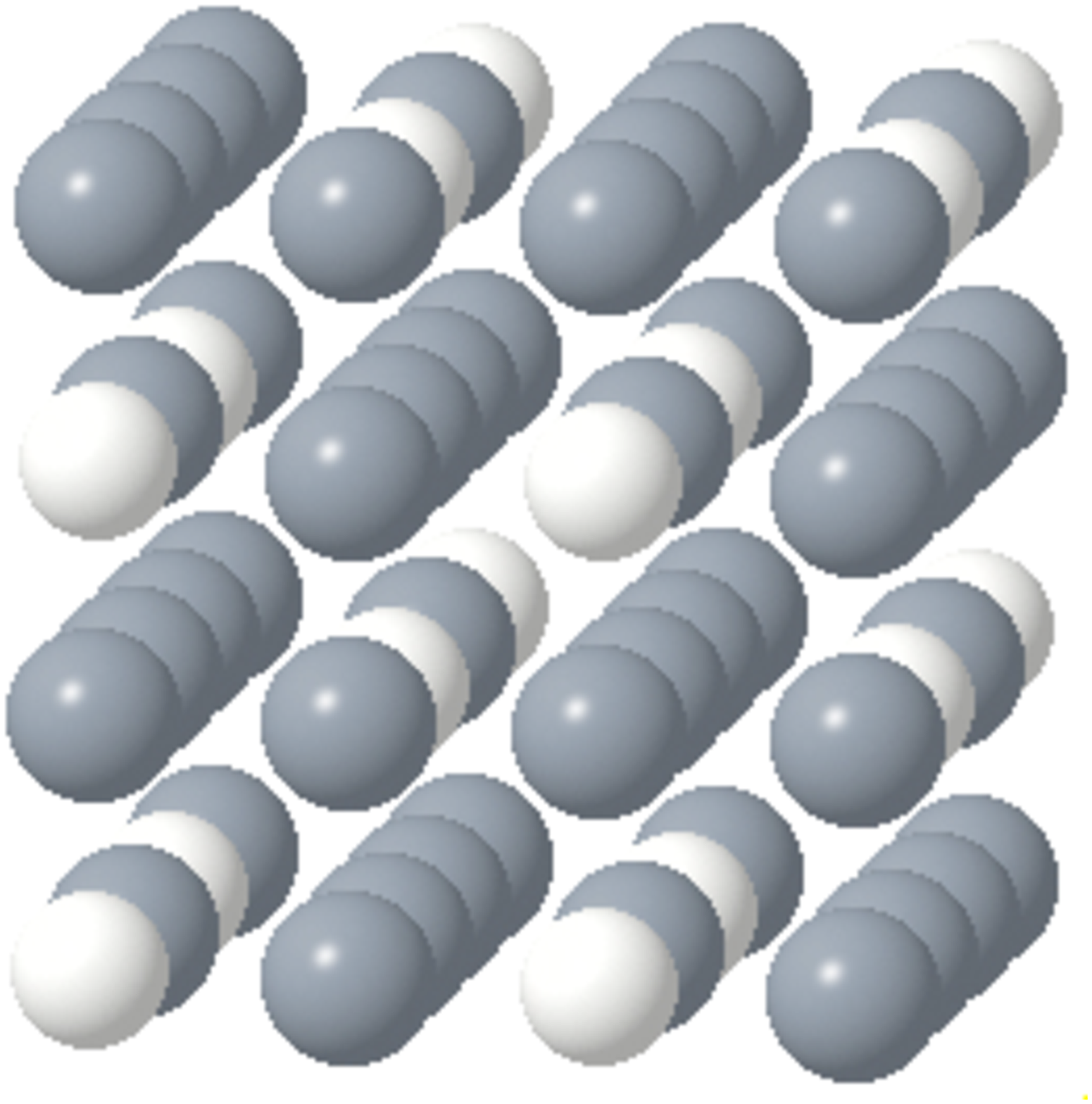}}
 \subfigure[]{
  \label{fig:subfig:crystdavide}
  \includegraphics[width=0.40\columnwidth,angle=0,keepaspectratio=T]
  {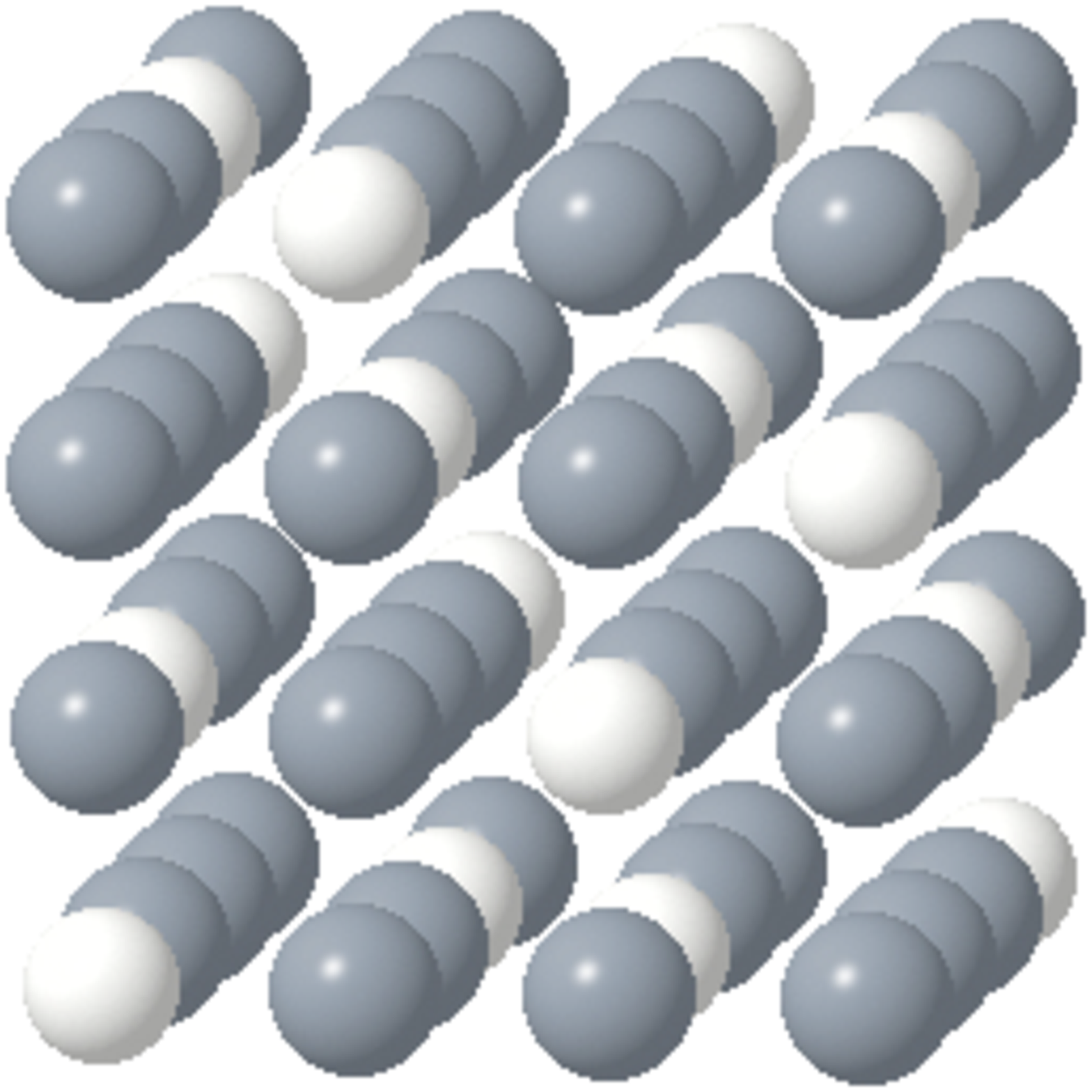}}
 \caption{Unit cells of the crystals at $T=0$.
 The black spheres represent an occupied position on the
 lattice and the white ones represent a vacancy.
 In both cases the density is $\rho=0.75$.
 It transpires that the crystal (a) is more stable (at low finite
 temperature) than (b), and therefore only (a) is observed by
 cooling in the simulation.}
\end{center}
\end{figure}

For $c_a=4$ and $c_r=3$ there are two
crystals with the highest energy $0.75$, made up of the couples
 $({}^{\phantom{0}}_1f_4^2,{}^{\phantom{0}}_0f_6^3)$,
$({}^{\phantom{0}}_1f_4^1,{}^{\phantom{0}}_0f_6^3)$, which
are represented in Figure \ref{fig:subfig:crysthector}
and \ref{fig:subfig:crystdavide}, respectively.
The point $\mu=-3\epsilon$
is still a multiphase point, but now \emph{four}
 particle-centered fragments
 $({}^{\phantom{0}}_1f_3^0,{}^{\phantom{0}}_1f_3^1,$
 ${}^{\phantom{0}}_1f_4^1,{}^{\phantom{0}}_1f_4^2)$
are degenerate. For this reason, here the multiphase point is much
more degenerate.

\begin{figure}[!htp]
\begin{center}
  \begin{overpic}[width=0.75\columnwidth,angle=-90]
   {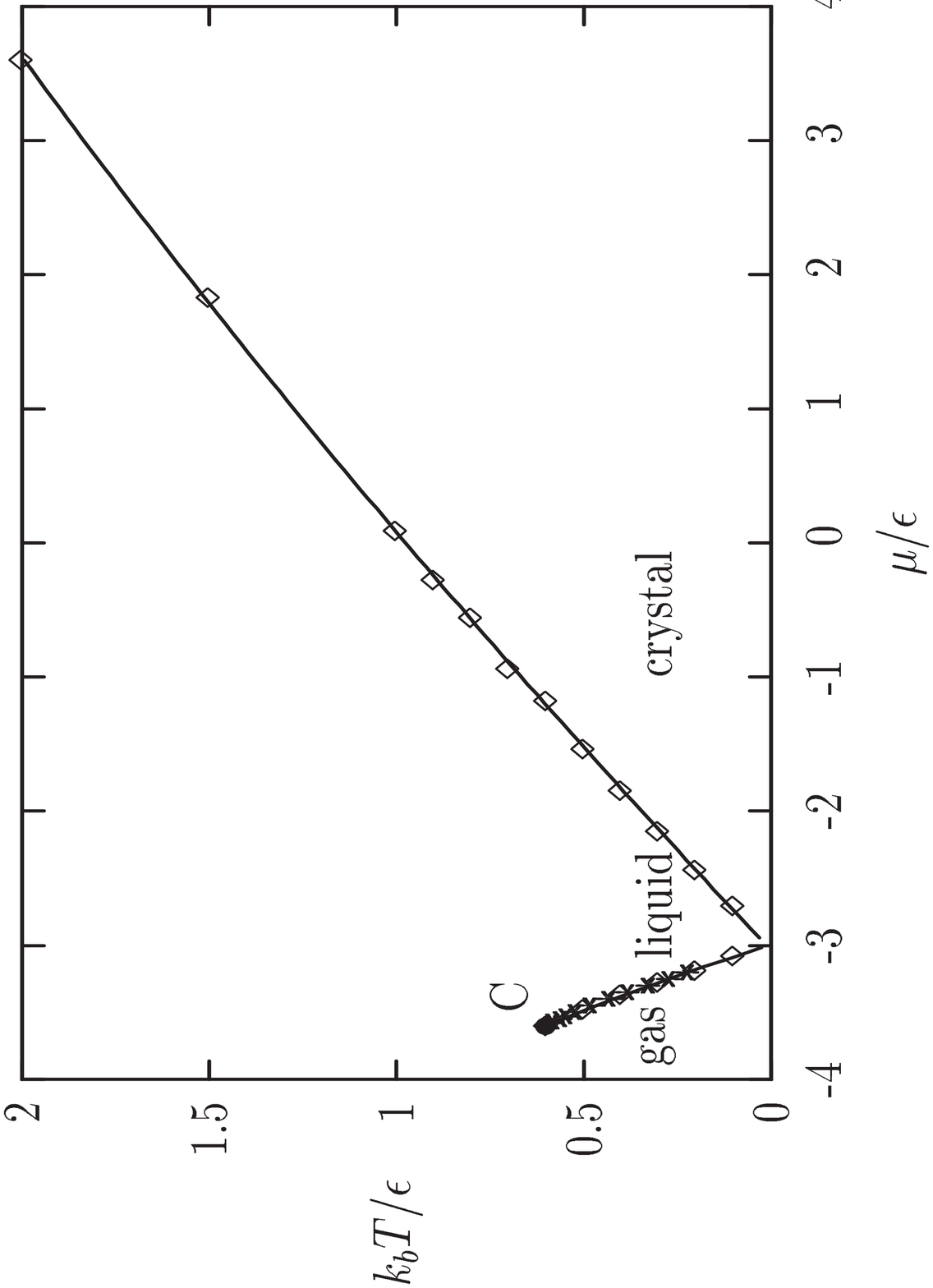}
   \put(17,65){
    \includegraphics[width=0.30\columnwidth,angle=-90]{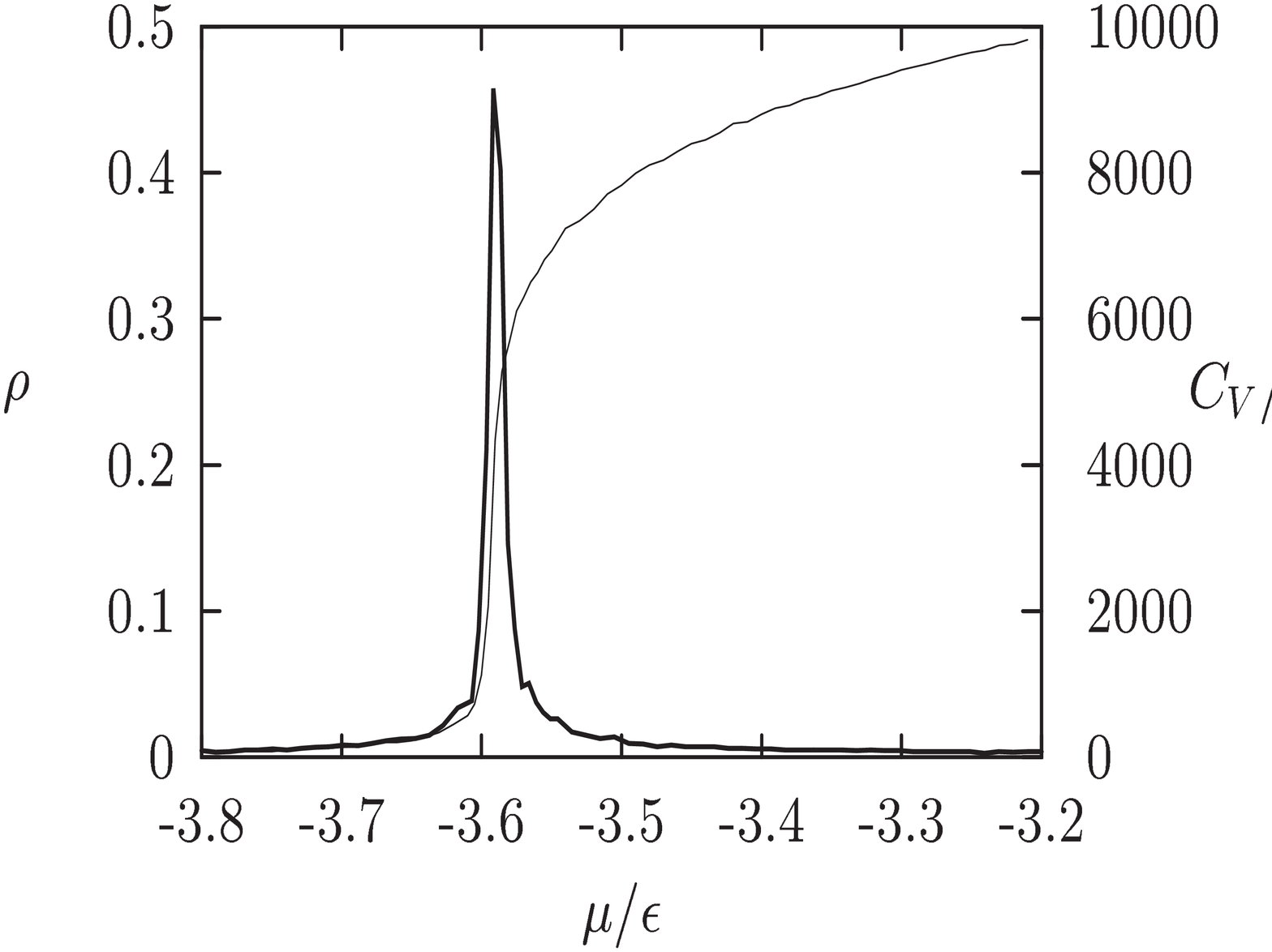}}
  \end{overpic}
 \caption{Monte Carlo Grand ensemble simulation for the case
 $c_r=4$, $c_a=3$.
 Graph of the temperature $(\beta\epsilon)^{-1}$ versus
 chemical potential ($\mu/\epsilon$).
 The ($\ast$) are the results of the simulation lowering
 the temperature. The ($\diamond$) are the results of the simulation
 increasing the chemical potential.
 The critical point is labelled C.
 The inset shows an isothermal scan of density and heat capacity
 for $k_bT=0.6\epsilon$.}
 \label{fig:phasedia2}
\end{center}
\end{figure}

\begin{figure}[!htp]
\begin{center}
  \begin{overpic}[width=0.75\columnwidth,angle=-90]
   {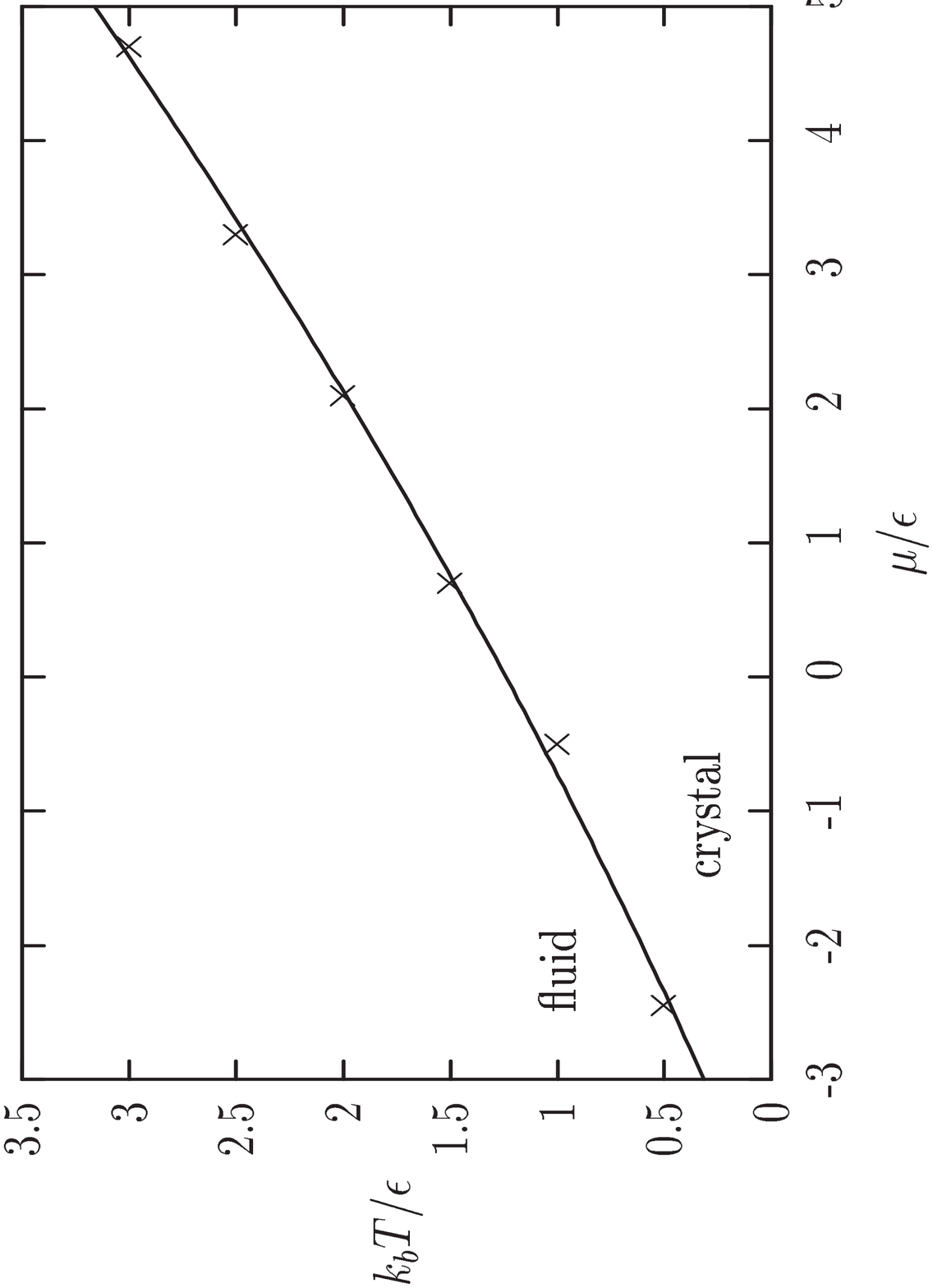}
   \put(16,65){
    \includegraphics[width=0.30\columnwidth,angle=-90]{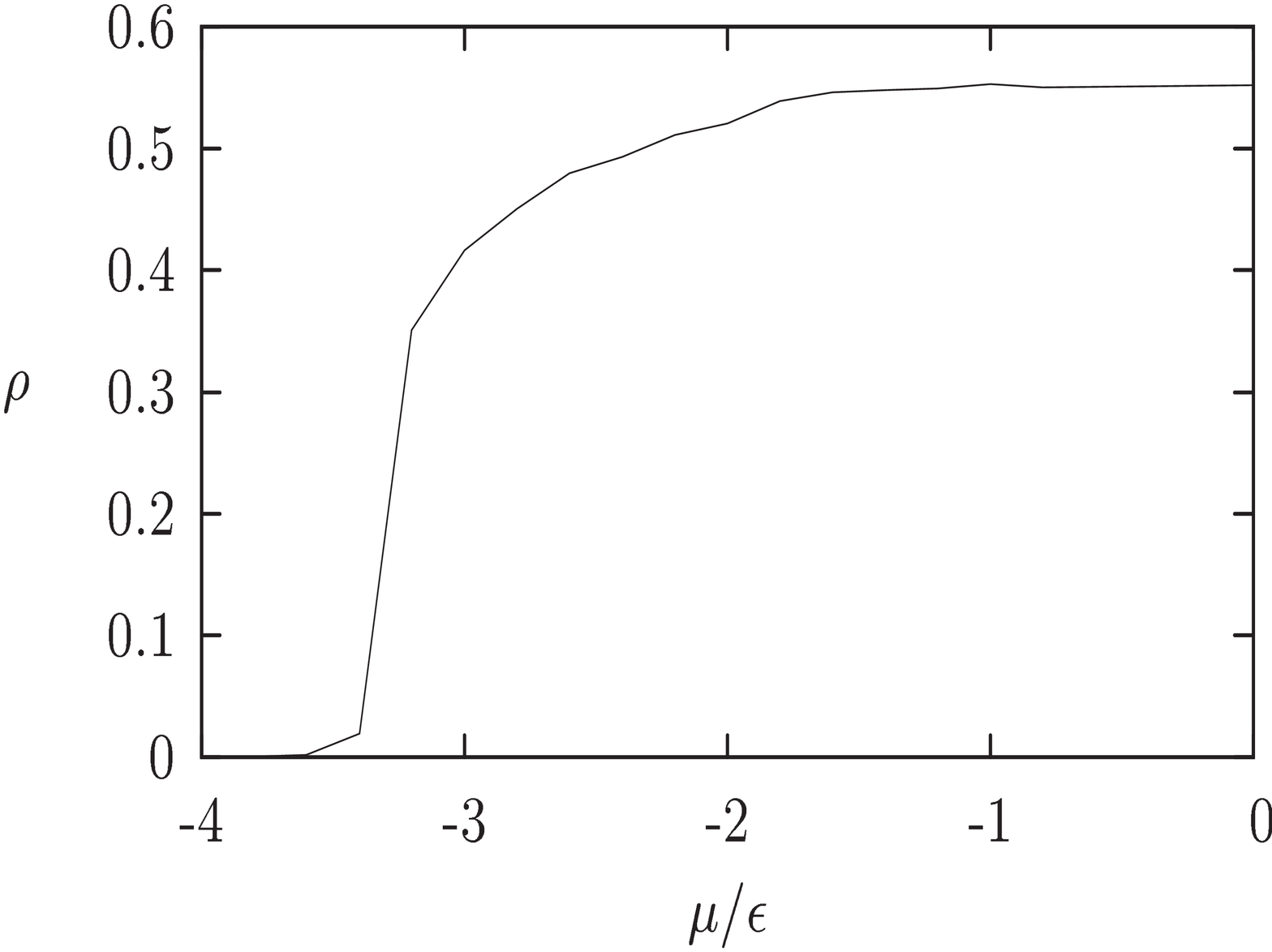}}
   \put(55,41){
    \includegraphics[width=0.28\columnwidth,angle=-90]{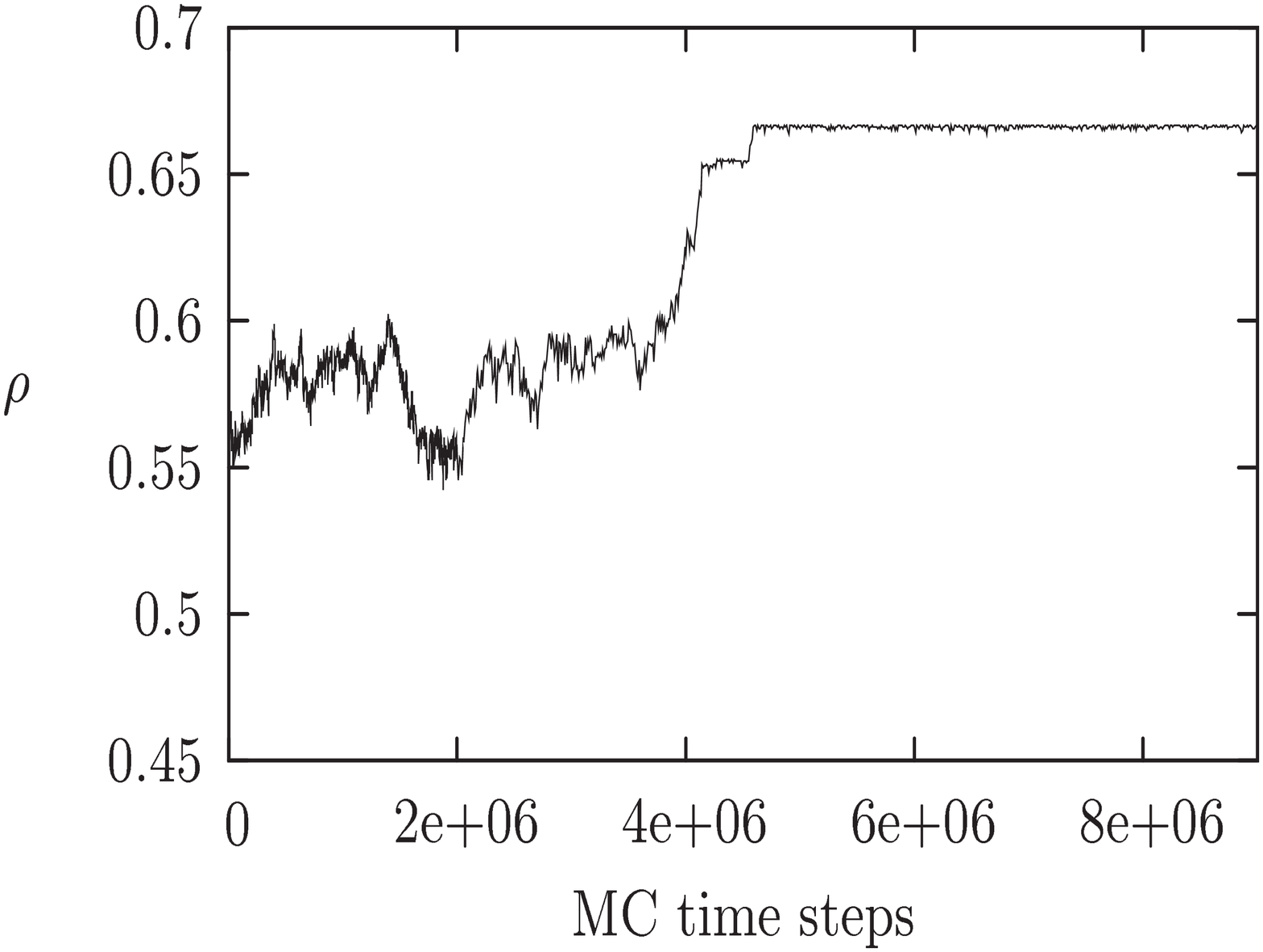}}
  \end{overpic}
 \caption{
 Monte Carlo Grand ensemble simulation
 for the case $c_r=3$, $c_a=3$.
 Graph of the temperature $(\beta\epsilon)^{-1}$ versus
 chemical potential ($\mu/\epsilon$).
 Upper inset: density \textit{vs}
 chemical potential graph for $k_bT=0.5\epsilon$.
 Lower inset: long time behavior for the point at $\mu=-2.4\epsilon$.}
 \label{fig:phasedia1}
\end{center}
\end{figure}

Besides determining the low temperature limit of the models, such
considerations allow us to estimate, and thereby control, as a
function of $c_a$ and $c_r$, the number of low temperature states
where attraction is dominant. This allows us ultimately to fix the
triple point of the system.

We present examples of Grand Canonical simulations ($L=12$,
$<10^6 MCS$). (We remark that we have as yet attempted no systematic exploration
of dynamical arrest except the expected transient
arrest phenomena of the underlying model \cite{birolimezard}).

Phase diagrams for $c_r=4$, $c_a=3$ and $c_r=3$, $c_a=3$
are shown in Figure \ref{fig:phasedia2} and
\ref{fig:phasedia1}, respectively.

The first is representative of a conventional gas-liquid-crystal
phase diagram, and the usual range of attractive interactions.
Thus the simulations readily reproduce the zero temperature states
of gas and crystal, and there is a significant range where the
liquid state is present. In the inset to Figure
\ref{fig:phasedia2} we show the density and heat capacity in an
isothermal cut of the phase-diagram, just above the gas-liquid
critical point. Such isotherms, as the temperature decreases from
this point, exhibit an increasing peak height of the heat
capacity, rising towards the critical point. The liquid-crystal
phase boundary is a conventional first-order phase-transition,
that asymptotes towards the fluid-repulsive crystal transition, as
expected. Simulations near the (low but finite $T$)
triple point become quite slow.

The second phase diagram is chosen to show how the model
represents a short range potential.
The simulations are still for quite small systems.
However, the gas-liquid branch, if it exists, is so short to be almost
unidentifiable. This would be the case, for example, of a
continuum square-well potential when the well width is less the
ten percent of the core size.

Furthermore, it is intriguing to note the numerous other
indications of a complex equilibrium, arrest, and kinetics
picture, as expected for the equivalent continuum picture. In
particular, there does appear to be a dense fluid-like state that
is long lived, as well as a string-like low density fluid that may
credibly identified as a gel, both expected. (In the upper inset to
the figure we show an isothermal scan in which the system makes an
apparent transition to a dense liquid-like phase, whereas in the
lower inset we show that this state degenerates to a crystal at very
long times)\cite{crystallization,foffipinidawson}.

In summary, we have shown that a very simple Ising lattice model,
with only nearest neighbor interactions, is capable of supporting
all of the characteristic phenomena of particles with repulsive
and attractive interactions; gas, liquid, crystals,
'glass-transition' phenomena of various kinds, and that variation
of parameters provides a mechanism to mimic the range of the
potential into important regimes that have yet to be understood.

Many of the outstanding problems of interest such as metastability
or buried critical points, arrest phenomena, and near-criticality
and gellation are expected to be quite well-described on lattices.
Crystallization nucleii and to some degree their growth are the
most dubious aspects of a lattice description, but even there many
of the most difficult and subtle unexplained processes currently
involve some element of near-criticality \cite{criticality}, something
which is well-dealt with in the lattice, and is difficult to deal
with in a useful manner otherwise.

Besides having a certain elegance in being a simplest model known
to exhibit all of the relevant phenomena, it seems likely that
this model has considerable relevance to important practical
issues.

\acknowledgements{
   We acknowledge with pleasure interactions with G. Biroli,
    S. Franz, G. Foffi, M. M\'ezard, M. Sellitto,
    F. Sciortino, P. Tartaglia and E. Zaccarelli.
The work is supported by DASM.  }

\bibliography{../bibtex/paper}

\end{document}